\documentclass[aps,prl,reprint,groupedaddress]{revtex4-1}
\usepackage{amsmath}
\usepackage{graphicx}  
\usepackage{dcolumn}   

\newcommand{\react}[1]{\stackrel{#1}{\longrightarrow}}
\newcommand{\ave}[1]{\langle #1\rangle}
\newcommand{\eq}[1]{Eq.\,\ref{#1}}
\newcommand{\eqs}[2]{Eq.\,(\ref{#1},\ref{#2})}

\begin{document}


\title{Applications of Little's Law to stochastic models of gene 
expression}


\author{Vlad Elgart}
\email{elgart@vt.edu}
\author{Tao Jia}
\email{tjia@vt.edu}
\author{Rahul V. Kulkarni}
\email{kulkarni@vt.edu}
\affiliation{Department of Physics, Virginia Polytechnic Institute and State University}


\date{\today}

\begin{abstract}
The intrinsic stochasticity of gene expression can lead to large
variations in protein levels across a population of cells.  To explain
this variability, different sources of mRNA fluctuations ('Poisson'
and 'Telegraph' processes) have been proposed in stochastic models of
gene expression.  Both Poisson and Telegraph scenario models explain
experimental observations of noise in protein levels in terms of
'bursts' of protein expression. Correspondingly, there is considerable
interest in establishing relations between burst and steady-state
protein distributions for general stochastic models of gene
expression.  In this work, we address this issue by considering a
mapping between stochastic models of gene expression and problems of
interest in queueing theory. By applying a general theorem from
queueing theory, Little's Law, we derive exact relations which connect
burst and steady-state distribution means for models with {\em
arbitrary} waiting-time distributions for arrival and degradation of
mRNAs and proteins. The derived relations have implications for
approaches to quantify the degree of transcriptional bursting and
hence to discriminate between different sources of intrinsic noise in
gene expression. To illustrate this, we consider a model for
regulation of protein expression bursts by small RNAs. For a broad
range of parameters, we derive analytical expressions (validated by
stochastic simulations) for the mean protein levels as the levels of
regulatory small RNAs are varied.  The results obtained show that the
degree of transcriptional bursting can, in principle, be determined
from changes in mean steady-state protein levels for general
stochastic models of gene expression.
\end{abstract}


\maketitle

\section{Introduction}

The intrinsic stochasticity of biochemical reactions involved in gene
expression can give rise to large variations in protein levels across
an isogenic population of cells \cite{kaern05, raj08}. Variations in
protein levels, in turn, can give rise to phenotypic heterogeneity and
non-genetic individuality in a population of cells \cite{avery06}.
The potential benefits of such phenotypic heterogeneity have been
discussed for diverse systems \cite{fraser09}. Correspondingly there
has been considerable effort focusing on uncovering molecular
mechanisms which drive gene expression 'noise' as a source of
phenotypic heterogeneity.

High variablity in protein levels has generally been attributed to
fluctuations in mRNA synthesis \cite{paulsson05,bareven06,newman06}.
To elucidate the source of mRNA fluctuations, two distinct models have
been proposed \cite{kaufmann07}. In one case, mRNA synthesis is
modeled as a Poisson process (Poisson scenario) and the high
variability in protein levels is related to low abundance and
infrequent synthesis of mRNAs \cite{thattai01}.  In the other case,
fluctuations are primarily driven by the slow kinetics of promoter
switching between active and inactive states (Telegraph scenario) with
mRNA synthesis occuring only during the active stage
\cite{raser05,paulsson05}. Recent work has further generalized these
models for gene expression to include the effects of processes that
can give rise to 'gestation' and 'senescence' periods for mRNA birth
and decay \cite{pedraza08}.  These terms derive from the observation
that both creation and degradation of cellular macromolecules
(mRNAs/proteins) often involve multiple biochemical
steps. Correspondingly, the waiting-time distributions for these
processes are more general than simple exponential distributions which
are characteristic of single-step Poisson processes.  Since the
observed noise in protein levels can include contributions from
different sources (e.g. transcriptional bursting as well as gestation
and senescence) most single-cell measurements of steady-state protein
distributions cannot be used to determine the source of fluctuations
in mRNA synthesis.

Recent studies have determined the variation of noise in protein
expression as a function of mean protein abundance for several genes
\cite{bareven06, newman06}. The observed scaling relationship is
consistent with both the Poisson and Telegraph scenario models in the
limit that protein production occurs in infrequent random bursts
\cite{bareven06, golding05}.  Furthermore, advances in single-molecule
techniques have led to studies monitoring real-time synthesis of
proteins in single cells \cite{yu06,cai06}. Protein expression was
indeed seen to occur in random bursts with a mean separation between
bursts that is large compared to typical mRNA lifetimes
\cite{yu06,cai06}.  Given these features, it is of interest to
consider whether quantification of protein burst distributions can
discriminate between the two scenarios for mRNA synthesis.  For the
Poisson scenario, the observed burst is a consequence of translation
from a single mRNA, whereas for the Telegraph scenario, it is produced
from a random burst of mRNAs synthesized when the promoter is in the
active state. While the underlying mRNA burst distributions can thus
be distinct for the two scenarios, it can be shown that observations
of protein burst distribution do not uniquely identify the underlying
mRNA burst distribution \cite{ingram08}.

Given that protein steady-state and burst distributions cannot
discriminate between the Poisson and Telegraph scenarios, it has been
argued that dynamic measurements of the number of mRNAs in single
cells are needed. Such methods have indeed been developed in recent
years \cite{golding05,raj08,larson09}, and have been used to quantify
the degree of transcriptional bursting. In this context, it would be
of interest to derive equations relating burst and steady-state
distribution means for both mRNAs and proteins. Such relations can
provide useful checks for experimental approaches for measuring
mRNA/protein burst distribution means. Furthermore they can also
suggest alternative approaches which allow inference of the underlying
mRNA burst distribution. This work focuses on deriving such relations
between the means of mRNA/protein burst and steady-state distributions
and exploring their consequences for approaches to quantify the degree
of transcriptional bursting.

In this paper, we consider a mapping between general stochastic models
of gene expression \cite{pedraza08} and problems of interest in
queueing theory. By applying a general theorem from queueing theory,
Little's Law, we derive exact relations connecting mRNA/protein burst
and steady-state distribution means for stochastic models of gene
expression with {\em arbitrary} waiting-time distributions for arrival
and degradation of mRNAs and proteins. Furthermore the derived
relations can be used to show how mRNA burst distributions can be
inferred from measurements of {\em mean} protein levels by introducing
an additional interaction in the reaction scheme. Specifically, we
consider a reaction scheme that includes interaction between mRNAs and
regulatory genes called small RNAs. In bacteria, small RNAs have been
studied extensively in recent years \cite{waters09} in part due to the
critical roles they play in cellular post-transcriptional regulation
in response to environmental changes. The results derived in this
work, besides the potential applications for quantifying the degree of 
transcriptional bursting, also provide insight into
small-RNA based regulation for specific parameter ranges.

\section{Model and Results}
\subsection{Connecting burst and steady-state means}

We begin by considering the minimal reaction scheme for translation from mRNAs 
\begin{equation}
        {M}\react{k_p}{M+P};\quad M\react{\mu_m}\emptyset;
        \quad P\react{\mu_p}\emptyset; \label{scheme-min}
\end{equation}
A single burst corresponds to proteins produced from the underlying
mRNA burst distribution until decay of the last mRNA. For the
Poisson process, mRNA transcription occurs with constant probability
per unit time $k_m$. On the other hand, for the Telegraph process,
mRNA transcription occurs with constant rate $k_m$ only when the DNA
is in the active(ON) state; once it transitions from the ON state to
the inactive OFF state (with rate $\alpha$) no mRNA transcription
can occur until it transitions back from the OFF state to the ON state
(with rate $\beta$). Note that in the limit $\beta << \alpha$,
mRNAs will be produced in infrequent bursts. Thus the Poisson process
gives rise to only protein bursts, whereas the Telegraph process gives
rise to both mRNA and protein bursts. It is interesting to note that
the mRNA burst distribution for both the Poisson and Telegraph
scenarios can be represented by the conditional geometric
distribution; specifically by considering bursts conditional on
production of at least one mRNA \cite{ingram08}. This can be
understood as follows: the mRNA burst distribution is the number of
mRNAs produced in the active state before transition to the inactive
state. Let us take the initial condition to correspond to the
transition from the inactive (OFF) state to the active (ON) state
i.e.\ at $t=0$ the DNA has just transitioned to the ON state. Now the
next reaction that can occur either results in the production of a
mRNA (with rate $k_m$) or a transtion to the OFF state (with rate
$\alpha$). The probablity of the next reaction being protein
production is $\frac{k_m}{k_m + \alpha}$ whereas the probability that
it is a transition to the OFF state is $\frac{\alpha}{k_m + \alpha}$. If 
we set $p = \frac{\alpha}{k_m + \alpha}$, the number of
mRNAs produced ($m$) before the transtion to the OFF state,
{\em conditional} on the production of atleast 1 mRNA, is given by
\begin{align}
	\pi_{\mathrm m}(m) & = (1-p)^{m-1}p,\quad m\ge 1,\label{geom}\\
	\pi_{\mathrm m}(0) & = 0,\label{cond}
\end{align}
For the conditional geometric distribution given above, the mean is 
given by 
\begin{equation}
m_b = \frac{1}{p}  \label{mean-cond-geom}
\end{equation}

For the Poisson scenario ($p=1$), a single mRNA is produced per burst, which
corresponds to the conditional geometric distribution with mean
$m_{b} = 1$.  The Telegraph scenario also gives rise to a conditional
geometric distribution for mRNA bursts, but with mean $m_{b} > 1$.
Thus, determination of the degree of transcriptional bursting
($m_{b}$) can discriminate between the Poisson and Telegraph scenarios
for intrinsic noise in gene expression.

The general model for gene expression that we analyze is as
follows. Bursts of protein expression result due to translation from
the underlying mRNA burst, which has a conditional geometric
distribution with mean $m_{b}$. The number of proteins produced from
different mRNAs are taken to be independent random variables. The
decay time for mRNAs and proteins is assumed to be drawn from
arbitrary waiting-time distributions with means $\tau_m$ and $\tau_p$
respectively. Likewise, the waiting-time distribution between
consecutive bursts is a random variable drawn from an arbitrary
distribution with mean $\tau_b$. Correspondingly, the average arrival
rate for bursts is given by $k_{b} = \frac{1}{\tau_b}$. Since the
waiting-time distributions are arbitrary, effects due to gestation and
sensecence of mRNAs and proteins \cite{pedraza08} are included.  For
this setup, we will derive analytical relations which can be used to
determine $m_b$ and thereby to quantify the degree of transcriptional
bursting.

We begin with the observation that the processes considered in the
above model have exact analogs in problems of interest in queueing
theory.  For example, the creation of proteins corresponds to the
arrival of customers in queueing models \cite{little}. On the
other hand, the service-time distribution corresponds to the
waiting-time for the customer to depart the system, making it the
analog of the waiting-time distribution for degradation of
proteins. Given that degradation of each mRNA/protein is independent
of other mRNAs/proteins in the system, the mapping corresponds to
queueing systems with infinite servers. This can be seen as follows.
In infinite server queues, since the number of servers is unlimited,
each customer is associated with a server immeduately upon
arrival. This effectively implies that each customer is served
independently of the others, which for the gene expression model is
equivalent to the assumption that mRNAs/proteins are degraded
independently.

A general theorem from queueing theory, Little's Law \cite{little},
states that the average number of customers in the system ($L$), the
mean arrival rate ($\lambda$) and the mean waiting time of a customer
in the system ($W$) are related by $L = \lambda W$.  The
remarkable feature of Little's Law is that it holds regardless of the
specific forms of the arrival and departure processes. When applied to
stochastic gene expression models, this implies that the processes
leading to mRNA/protein can be arbitrary, e.g. including gestation and senescence 
effects.

We now apply Little's Law to derive an equation relating mRNA burst
and steady-state distribution means.  The arrival rate of mRNA bursts
is driven by an arbitrary stochastic process with
average arrival rate $k_b$. The decay process of mRNA is also assumed
to be driven by an arbitrary stochastic process with average decay
time $\tau_m$.  Employing Little's Law \cite{little}, we obtain a relation 
between the mean mRNA burst size $m_b$ and the average number of mRNAs in the 
steady state:
\begin{align}
	\ave{m} = \lambda \tau_m,\label{Little}
\end{align}
where $\lambda$ is average arrival rate of the mRNAs, which
is given by
\begin{align}
	\lambda = m_b k_b \label{arrival}
\end{align}
Hence, we derive that the steady-state distribution mean for mRNAs is
related to the mean mRNA burst size by 
\begin{align}
	\ave{m} = m_b k_b \tau_m \label{mean-mRNA}
\end{align}

Both the mean steady-state mRNA levels ($\ave{m}$) and the mean mRNA
lifetime ($\tau_{m}$) can be determined experimentally using standard
procedures. \eq{mean-mRNA} implies that the degree of
transcriptional bursting can then be determined by estimating the mean
burst arrival rate ($k_{b}$), which can be done using single-molecule
approaches. Such a procedure was used in Ref. \cite{cai06} to estimate
the degree of transcriptional bursting, with the assumption of
constant mRNA arrival rates and decay rates. \eq{mean-mRNA} indicates
that, even if this is not the case and arbitrary gestation and
sensescence periods are considered, the above procedure remains a
valid approach to determine the degree of transcriptional bursting
$m_{b}$. Alternatively, since the above relation is valid for arbitrary 
stochastic processes governing mRNA arrival and decay, it can serve as 
a useful consistency check for different experimental approaches for 
quantifying mRNA burst and steady-state distributions.

Using Little's Law we can also relate the steady-state protein distribution 
mean to the burst mean following similar logic.  Since the
average arrival rate of proteins is given by $\ave{m} k_p$, we derive
\begin{align}
	P_s  = \ave{m}k_p\tau_p,\label{mean-proteins}
\end{align}
where $k_p$ and $\tau_p^{-1}$ are average synthesis and decay rates of
the proteins.  The above equation can be recast in terms
of the mean number of proteins produced in a single burst $P_{b}$
(which is related to the mRNA burst distribution mean by $P_{b} =
m_{b} k_{p} \tau_m$).  Since the mean arrival rate of proteins is
given by $k_{b}P_{b}$, we have
\begin{equation}
P_{s} = k_{b} P_{b} \tau_{p} \label{protein-little}
\end{equation}
It is noteworthy that this simple relation is valid for arbitrary
gestation and senescence waiting-time distributions.  It establishes
that the mean steady-state protein level only depends on the average
protein arrival and degradation rates and is independent of the higher
moments of the corresponding waiting-time distributions.  Thus, it
explains the observation in Ref \cite{pedraza08} that gestation and
senescence do not affect the average susceptibility to changes in
parameters.

Another important consequence of \eq{protein-little} is that processes
that alter the burst distribution mean without affecting protein
degradation times or burst arrival times will produce a proportionate
change in the steady-state distribution mean.  Thus, regulatory
interactions which are sensitive to the degree of transcriptional
bursting and alter protein burst distributions will produce
proportionate changes in protein steady-state distribution
means. This, in turn, suggests the possibility of obtaining signatures
of transcriptional bursting by observing {\em changes} in steady-state
protein distribution means upon regulation. To explore this
possibility, let us consider how regulation by small RNAs modulates
protein burst distributions.

\subsection{Regulation by small RNAs}

\begin{figure}[tb]
\begin{center}
\resizebox{6cm}{!}{\includegraphics{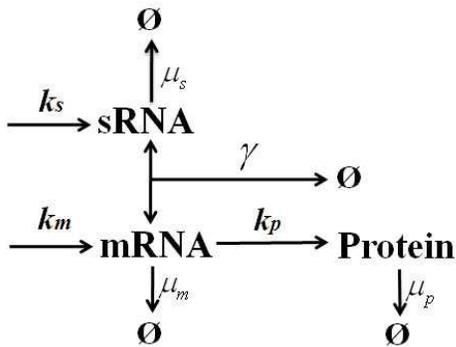}}
\caption{The kinetic scheme for regulation of protein production by small RNAs with coupled degradation rate $\gamma$. \label{action}}
\end{center}
\end{figure}

We consider regulation by small RNAs (sRNAs) based on a coarse-grained
model (Fig. \ref{action}) studied previously \cite{levine07,mehta08,mitarai07}
which applies to sRNAs that regulate mRNA targets stoichiometrically
due to coupled degradation \cite{masse03}. Synthesis of sRNAs is taken
to be a Poisson process with constant rate $k_{s}$ and the sRNA
degradation rate is also taken as constant ($\mu_{s}$) in the
following analysis.  The parameter $\gamma$ controls mutual
degradation of mRNAs interacting with sRNAs. 
As in the previous section, mRNAs are created in bursts, with the average 
rate of arrival for bursts given by $k_b$. If $k_{b}\mu_s << 1$, i.e.
if the sRNA lifetime is
small compared to the mean arrival time between bursts, the distribution of sRNAs {\em prior} to a mRNA burst can
be approximated by the steady-state distribution of sRNAs in the
absence of mRNAs. Given this approximation, we wish to derive
expressions for the protein burst distribution in the presence of
sRNAs. This is, in general, analytically intractable. However by
employing further approximations which are valid for a range of
parameters we can obtain analytical expressions for the burst
distribution. Specifically, we assume that synthesis of new sRNAs {\em
during} a burst can be ignored, i.e. no new sRNAs are created in the 
time interval between mRNA creation and decay. Furthermore, we consider $\gamma \tau_m \gg 1$ such that mRNA degradation in the presence of sRNAs is assumed
to occur due to mutual degradation with a sRNA rather than natural
decay with average rate $\mu_m = \frac{1}{\tau_m}$. Given that these
approximations are valid, a simple analytic expression for the mean
regulated protein levels can be obtained as a function of mean sRNA
levels as shown below.

\noindent\\

As indicated in the reaction scheme in Fig.\ref{action}, a pair of
molecules of mRNA and sRNA can combine and be degraded rapidly with
rate $\gamma$. We first consider the limit $\gamma\rightarrow\infty$.
In this case, regulation by sRNA results in an instantaneous
modification of the distribution of mRNAs just after the burst.
The mRNA burst distribution prior to interaction with sRNAs is given by
$\pi_{\mathrm m}(m)$ (\eq{geom}).  The modified mRNA burst distribution
after interaction with sRNAs ($\tilde \pi_{\mathrm m}(m)$) is given by
\begin{align}
	\tilde \pi_{\mathrm m}(m) & = \sum_{n=0}^\infty \rho(n) \pi_{\mathrm m}(m+n),\quad m\ge 1,\label{geom-reg}
\end{align}
where $\rho(n)$ is the probability of finding $n$ sRNA molecules at
the time of burst. Any burst of $m$ mRNA molecules instantly becomes
an effective burst of $m - n $ mRNA molecules (for $ m > n$) due to
coupled degradation with $n$ sRNAs.  If $m\le n$, the mRNA burst after
the regulation will be effectively an `empty' burst. The probability
of an empty burst is given by
\begin{align}
	\tilde \pi_{\mathrm m}(0) & = 1-\sum_{m=1}^\infty\tilde \pi_{\mathrm m}(m).\label{cond-reg}
\end{align}

\par

Since the unregulated mRNA burst distribution is geometric (with
parameter $p$, say), we derive
\begin{eqnarray}
\tilde \pi_{\mathrm m}(m) &=& \pi_{\mathrm m}(m)\sum_{n=0}^\infty (1-p)^n \rho(n)  \nonumber \\
&=& G(1-p) \pi_{\mathrm m}(m),\quad m\ge 1,\label{geom-reg-2}
\end{eqnarray}

where $G(1-p)$ is the generating function of sRNA probability
distribution $\rho(n)$, evaluated at the point $1-p$. Using
\eq{cond-reg} we derive
\begin{align}
	\tilde \pi_{\mathrm m}(0) & = 1 - G(1-p),\label{cond-reg-2}
\end{align}

The regulated mRNA burst distribution is thus a conditional geometric
distribution as in the unregulated case, but with modified average
arrival rate $\tilde k_m = G(1-p) k_m$. This is because $1 - \tilde
\pi_{\mathrm m}(0) = G(1-p)$ is the probability that the {\em regulated} burst
results in atleast 1 mRNA. Therefore, the average number of mRNAs in
the steady state for the regulated case is given by (according to the
equation \eq{mean-mRNA})
\begin{align}
	\ave{\tilde m} = m_b \tilde k_m\tau_m = G(1-p)\ave{m}.\label{mean-mRNA-reg}
\end{align}
\par

We denote by $P_{b}(n_{s})$ the mean burst size for proteins in the
presence of sRNAs, where $n_{s} = k_{s}/\mu_{s}$.  Using the equations
\eqs{mean-proteins}{mean-mRNA-reg} for mRNA's steady state average, we
derive in the limit of fast coupled degradation
($\gamma\rightarrow\infty$)
\begin{align}
	P_s(n_s) = G(1-p)P_s(0).\label{mean-proteins-reg}
\end{align}
Taking the sRNA distribution prior to the burst ($\rho(n)$) to be a
Poisson distribution with mean $n_{s} = k_{s}/\mu_{s}$, and given that
the mean mRNA burst size is given by $m_b = 1/p$ (\eq{mean-cond-geom}), we 
derive
\begin{equation}
\frac{P_{b}(n_{s})}{P_{b}(0)} =  e^{-n_{s}/m_{b}} \label{burst-mean}
\end{equation}
Thus, if the burst mean ($P_{b}$) is determined along with $n_{s}$,
the above relation determines $m_{b}$ and hence the degree of
transcriptional bursting.  \eq{protein-little} further implies that
the ratio of protein steady-state means for regulated to unregulated
cases ($\frac{P_{s}(n_{s})}{P_{s}(0)}) $ is equal to the corresponding
ratio for the burst means in \eq{burst-mean}. This in turn implies
that the mean transcriptional burst size $m_{b}$ can be determined by
considering changes in {\em mean} steady-state protein
levels. Taken together, these results provide a novel procedure for
determining $m_{b}$.

\begin{figure}[t!]\vspace*{3pt}
\centering{\resizebox{6cm}{!}{\includegraphics{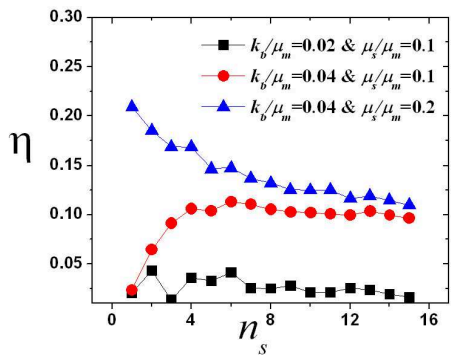}}\vspace*{-5.5pt}
\caption{ The relative error $\eta$ of the estimated $m_b$ value
(actual value $m_b = 10$) as a function of mean sRNA levels $n_s$. The
estimate from the flat portions of the curves is within $15\%$ error
for the parameters shown with $ \frac{\gamma}{\mu_m} = 20$. The
results shown are for the case of exponential waiting-time
distributions for mRNA arrival and decay, comparable relative errors
were obtained for more general waiting-time
distributions. \label{steady}} } \vspace*{-8pt}
\end{figure}

The proposed procedure has been computationally validated for a range
of parameters using stochastic simulations (Fig. 2).  Specifically, we
set up simulations based on the standard Gillespie algorithm
\cite{gillespie77} wherein the waiting-time for the next reaction is
drawn from an exponential distribution.  To consider effects such as
mRNA senescence, we model mRNA degradation as a multi-step process,
wherein the waiting-time distribution for each step is drawn from an
exponential distribution such that the degradation time for mRNAs
follows a Gamma distribution (see Appendix).  Similarly, mRNA arrival was 
simulated
as a multi-step process with gamma waiting-time distribution between
mRNA arrival bursts.  The output from the simulations is the mean
steady-state protein levels as a function of the mean sRNA levels
($n_{s}$), where the mean sRNA levels are varied by increasing the
sRNA creation rate $k_{s}$.  Provided that the system parameters are
consistent with the following constraints: $\gamma \gg \mu_m \gg
\mu_s$, simulations indicate that the transcriptional burst size
$m_{b}$ can be predicted with reasonable accuracy from the ratio of
measured protein steady-state means for regulated to unregulated cases
as discussed above. The errors in the estimate for $m_{b}$ using the
above procedure are related to the validity of the approximations made
and are discussed further in the Appendix. Provided that regulatory
small RNAs can be designed with parameters subject to the constraints
noted, the relative error in estimating $m_{b}$ is small and thus we
can determine the degree of transcriptional bursting and clearly
distinguish between the Poisson and Telegraph scenarios. The parameter
ranges for validity of the above analysis are accessible experimentally 
based on previous work, e.g. high values of $\gamma$ relative to the natural 
degradation rate $\mu_m$ are expected for the sRNA RhyB \cite{mitarai09},
and mRNA burst arrival rates which are small compared to the mRNA 
degradation rate have also been reported \cite{cai06}.
Finally, we note that it would be of interest to apply the preceding analysis 
to systems which show high degree of transcriptional bursting primarily 
arising from random activation and inactivation of the promoter state 
\cite{raj06}. In particular, it was observed \cite{raj06} that increasing 
concentrations of a transcriptional activator resulted in increasing the mean
burst size rather than affecting the burst frequency. Since the procedure 
proposed in this work is an independent approach to determine the burst mean, 
it would be of interest to further analyze the above system using the analysis
proposed in the current work.

\section{Summary and Conclusions}

In summary, we have considered a generalized model of gene expression
with bursty production of mRNAs and proteins. Since very different
stochastic processes can lead to steady-state distributions that are
experimentally indistinguishable, the degree of transcriptional
bursting cannot be inferred from steady-state protein
distributions. In light of this, it has been argued that determination
of transcriptional bursting requires dynamic measurements of mRNA
molecules in single cells \cite{golding05,pedraza08,raj09}. In this
work, we have derived exact relations connecting mRNA/protein burst
and steady-state distribution means which are valid for arbitrary
gestation and senescence waiting-time distributions. We further
analyzed how protein burst distributions are modified due to
regulation by small RNAs for a range of parameters. Our analysis 
computationally demonstrates an alternative procedure for quantifying
transcriptional bursting, which involves measurements of changes in
mean protein steady-state levels induced by interactions with small
RNAs. The strategy presented can also be applied to a broader classes
of biological networks whose analysis requires inference of internal
variables from observations at higher levels.  An alternative strategy
to direct measurements of internal variables is to discriminate
different possibilities for the internal variables by coupling to a
controlled external interaction.

\section{Appendix}

\subsection{Finite $\gamma$ corrections}

The analysis in the main text considered the limit
($\gamma\rightarrow\infty$) and we now consider corrections due to
finite $\gamma$ values.  Lets take a more detailed look at the protein
production process during the burst.  We denote the duration from the
beginning of the burst to the time when sRNA or mRNA number first
reaches zero as the first stage of the burst. If the mRNAs outnumber
the sRNAs, excess mRNAs will be left after the coupled degradation and
evolve accordingly.  We call the duration from this point to the time
when all mRNAs are degraded as stage two of the burst.  In the case
that $\gamma\rightarrow\infty$, the duration of stage one will be zero
and all proteins are produced in stage two of the burst. However, for
finite $\gamma$ value, one has to take into account proteins that have
been synthesized during stage one of the burst. \\

In order to estimate the amount of the proteins produced on average
from mRNAs that are degraded by sRNAs (stage one), we observe first
that the {\em minimal} degradation rate of a single mRNA in this
process is $\gamma$. This is because at least one sRNA should be
present to ensure coupled degradation.  Second, the total amount of
mRNAs in the originating burst is greater than or equal to number of
mRNAs degraded by sRNAs (since some mRNAs may decay naturally). \\

Hence, we can employ formula \eq{mean-mRNA} in order to estimate
contribution of the mRNAs decaying in coupled degradation process to
overall steady state level. The upper bound of this contribution is
given by
\begin{align}
	\ave{\delta m} \sim n_m \frac{k_m}{\gamma}.
\end{align}
Here we replaced the rate $\tau_m^{-1}$ in the \eq{mean-mRNA} by the
minimal rate $\gamma$ in order to estimate the upper limit. \\

Now we can use the expression \eq{mean-proteins} to get the upper
bound of the proteins produced on average from the mRNAs during the
coupled degradation process
\begin{align}
	\delta P_s(n_s) \sim \delta m k_p\tau_p \sim \left(n_m \frac{k_m}{\gamma}\right) k_p\tau_p.
\end{align}

\begin{figure}[tb]
\begin{center}
\resizebox{7cm}{!}{\includegraphics{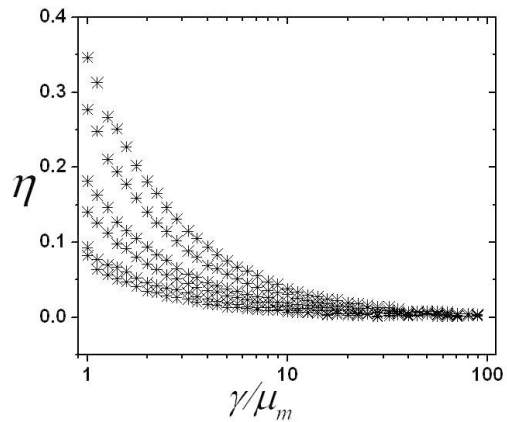}}
\caption{The relative error $\eta$ of the estimated $m_b$ derived from changes in  mean protein burst levels. For different $n_s$ and $m_b$ values, the error is negligible when $\gamma \ge 10\mu_m$. \label{gamma}}
\end{center}
\end{figure}\noindent

Hence, the overall ratio of regulated to unregulated mean steady state
levels of proteins is bounded as
\begin{align}
	\delta R \sim \frac{\delta P_s(n_s)}{P_s(0)} = \frac{1}{\tau_m\gamma},
\end{align}
which is independent of protein's synthesis rate $k_p$. Therefore, if
coupled degradation process is much faster than natural mRNA decay,
$\tau_m\gamma\gg 1$, we obtain $\delta R \rightarrow 0$ which is
validated by simulations. As we can see in Fig(\ref{gamma}), when
$\tau_m\gamma > 10$ the proteins produced during stage one of the
burst can be neglected and the result is almost the same as when
$\gamma \rightarrow \infty$.  Finally we note that recent studies
\cite{mitarai09} have shown that a well-studied bacterial small RNA
(RhyB) does induce rapid degradation of target mRNAs consistent with
the condition $\tau_m\gamma \gg 1$.

\subsection{Waiting-time distribution for multi-step processes}

Previous work \cite{pedraza08} on gestation and senescence
effects in mRNA/protein production and decay considered extensions of
the single-step Poisson process to multi-step processes. For the
simplest case, the corresponding waiting time distribution is a Gamma
distribution as derived below. Consider a multi-step process,
consisting of $n$ steps such that each step is completed with rate
$k$.  Let $T$ denote the random variable corresponding to the
waiting-time for the process to finish and let $T_{i}$ be the random
variable corresponding to the waiting-time for the $i^{th}$ step. Thus
we have $ T = \sum_i T_{i} $, i.e. $T$ is the sum of $n$ identical
independent random variables. Correspondingly the Laplace transform of
the probability distribution for $T$ (denoted by $F(s)$ say) is given
by the product of $n$ Laplace transforms of the exponential
distribution. The exponential waiting-time distribution for the
$i^{\mathrm th}$ step is given by $ke^{-k t}$ with corresponding
Laplace transform $\frac{k}{k + s}$. Correspondingly we have $F(s) =
(\frac{k}{k + s})^n$, and inverting the Laplace transform we obtain
that the waiting-time distribution for the multi-step process is given
by the Gamma distribution: $k (k t)^{n-1} \frac{ e^{- k t}}{(k-1)!}
$.

\bibliographystyle{apsrev4-1}
\bibliography{srna_burst}

\end{document}